\documentclass{PoS}

\title{Scaling and low energy constants in
  lattice QCD with $N_{\rm f}=2$ maximally twisted Wilson quarks}

\ShortTitle{Scaling and low energy constants in $N_{\rm f}=2$ Mtm-LQCD
  \hspace*{1.3cm} {\rm Roberto Frezzotti and Gregorio Herdoiza}}

\author{Petros Dimopoulos,~Roberto Frezzotti\thanks{Speaker.}~~and Gregorio Herdoiza$^*$\\
  Univ. and INFN of Rome Tor Vergata, Via della\ Ricerca Scientifica 1, I-00133, Rome, Italy\\
  E-mail: \email{\{dimopoulos,frezzotti,herdoiza\}@roma2.infn.it}}
\author{Carsten Urbach\\
  Theoretical Physics Division, Dept. of Mathematical Sciences,
  University of Liverpool \\ Liverpool L69 7ZL, UK \\
  E-mail: \email{carsten.urbach@physik.hu-berlin.de}}
\author{Urs Wenger\\
  Institute for Theoretical Physics, ETH Z{\"u}rich, CH-8093 Z{\"u}rich,
  Switzerland\\
  E-mail: \email{wenger@itp.phys.ethz.ch}}
\vspace{0.5cm}
\author{for the ETM Collaboration}
\vspace{0.5cm} \abstract{We report on the scaling of basic hadronic
  observables in lattice QCD with $N_{\rm f}=2$ maximally twisted
  Wilson dynamical quarks. We give preliminary results for some of the
  Gasser--Leutwyler low energy constants, the chiral condensate and
  the average mass of $u$ and $d$ quarks.}

\FullConference{The XXV International Symposium on Lattice Field Theory\\
                 July 30 - August 4 2007\\
                 Regensburg, Germany}

\usepackage{subfigure}
\begin{document}

\section{Introduction and setup}

In this contribution we report on the scaling of basic hadronic
observables and present preliminary results for some of the
Gasser--Leutwyler low energy constants in lattice QCD with $N_{\rm
f}=2$ dynamical quarks. We choose the lattice formulation with
tree-level improved gauge action and maximally twisted Wilson
quarks~\cite{ETM1}, which can be efficiently studied by means of
state-of-the-art simulation algorithms, such as the one we
adopted~\cite{UJSW}, and leads to physical observables free of O($a$)
cutoff effects~\cite{FR1}. In this way it is possible to study the
theory with pion masses down to about 300~{\rm MeV}, three different
lattice resolutions and spatial lattice sizes of 2--3 fm.  The main
parameters of the simulations employed for the present analysis are
summarized in table 1 of Ref.~\cite{CU-PROC}. For all ensembles we
have $m_{\rm PS}L \geq 3$, with the lowest values (namely 3.0, 3.3,
3.3 and 3.5) being obtained in the ensembles C6, B1, C1 and C5,
respectively.  For details about our lattice setup, the evaluation of
quark propagators and any undefined notations we refer to
Refs.~\cite{ETM1,CU-PROC}.

\subsection{Tuning to maximal twist}
\label{tuneMT}

The values of $\kappa$ in table 1 of Ref.~\cite{CU-PROC} result from
implementing maximal twist as discussed here. The formal definition of
maximal twist for Wilson quarks reads: $m_{\rm R}=0$ and $\mu_{\rm R}
= {\rm O}(a^0)$ for all lattice spacings $a$ as $a\to 0$, with
renormalized mass parameters
\begin{equation}
  \mu_{\rm R} = Z_\mu \mu = Z_P^{-1}\mu \; , \quad\quad
  m_{\rm R} = Z_{S^{\;0}}^{-1}(m_0 - m_{\rm crit}) 
  = Z_A Z_{P}^{-1} m_{\rm PCAC} \; .
\end{equation}  	
At the non-perturbative level any legitimate estimate of the critical
mass, $m_{\rm crit}$, is affected by terms of O($a\Lambda_{\rm
QCD}^2$) and O($a\mu\Lambda_{\rm QCD}$), which however do not
invalidate the definition of maximal twist. Following
Refs.~\cite{ShWu2,FMPR}, for each $\beta$ (and $\mu$) one can define
maximal twist by demanding\,\footnote{Here quark bilinears are written
in the unphysical quark basis ($\chi$, $\bar{\chi}$) where the Wilson
term has its standard form.}
\begin{equation}
  am_{\rm PCAC}(\beta,\mu) \equiv 
  \frac{a^4 \sum_{\bf x} \partial_0 \langle \bar{\chi}\gamma_0\gamma_5\tau^1\chi(x) 
    \bar{\chi}\gamma_5\tau^1\chi(0)  \rangle}
       {a^3\sum_{\bf x} \langle \bar{\chi}\gamma_5\tau^1\chi(x) 
	 \bar{\chi}\gamma_5\tau^1\chi(0) \rangle}\Big{|}_{\beta,\mu}   = 0 \, ,
       \label{mpcac0_1}
\end{equation}
at values of $x_0$ and $L$ so large that the (charged) one-pion state
dominates the correlators on the r.h.s. This prescription fixes the
lattice artifact O($a\Lambda_{\rm QCD}^2$) in the critical mass in
such a way that, provided $ \mu \gtrsim a^2\Lambda_{\rm QCD}^3 $, the
dominating (relative) cutoff effects left-over in physical
observables\,\footnote{Concerning the $\pi^0$-mass, in
Ref.~\cite{GCRtalk} it is argued that the possibly large O($a^2$)
artifact on this observable is merely due to the large value taken by
a (continuum) matrix element present in the Symanzik expansion of all
lattice correlators where the $\pi^0$ state contributes, rather than
to the presence of dimension six operators with large coefficients in
the Symanzik effective action. This implies that this cutoff effect
represents an exceptional, though important case.}  are expected to be
numerically as small as O($a^2\Lambda_{\rm QCD}^2$).  A detailed
analysis~\cite{FMPR} shows that such cutoff effects are actually
products of (two) terms of order $a\Lambda_{\rm QCD}$, $a\mu$
(negligible for small $\mu$) or $a^3\Lambda_{\rm QCD}^4\mu^{-1}$
(higher order in $a$, but enhanced for small $\mu$).

In practice, as we are interested in simulations with $\mu \gtrsim
\mu_{\rm LOW}$, in order to minimise the work for the tuning of
$\kappa$, we choose to impose the condition~(\ref{mpcac0_1}) only for
$\mu=\mu_{\rm LOW}$.  With this choice, in the region where $\mu <
\Lambda_{\rm QCD}$ we expect~\cite{FMPR,ETM1} the numerically
dominating cutoff effects on physical observables to be modulated by
factors of $\mu_{\rm LOW}/\mu$ as $\mu$ is varied (for instance one
can have contributions from terms of order $(a\Lambda_{\rm
QCD}\mu_{\rm LOW}/\mu)^2$).  With $\kappa$ fixed according to the
criterion described above, maximal twist is implemented properly,
provided
\begin{equation}
\mu \gtrsim \mu_{\rm LOW} \geq C a^2\Lambda_{\rm QCD}^3  \, ,
\label{THmuWIND}
\end{equation}
where the coefficient $C$ has to be learned from numerical experiment.
In our $N_{\rm f}=2$ setup,~$C~\sim~2$ (assuming $\Lambda_{\rm QCD} =
250$~{\rm MeV}) is essentially determined by the condition that MC
simulations exhibit no metastabilities (see e.g.\ Ref.~\cite{AS-REP},
sect.~5.5). At fixed $\mu_{\rm R}$ a smooth approach to the continuum
limit is to be expected if maximal twist has been realized at
(approximatively) the same value of $\mu_{\rm R \; LOW}$ in physical
units for all the considered lattice resolutions.

\begin{figure}[!h]
\begin{center}
\subfigure[]{\includegraphics[width=0.513\textwidth]{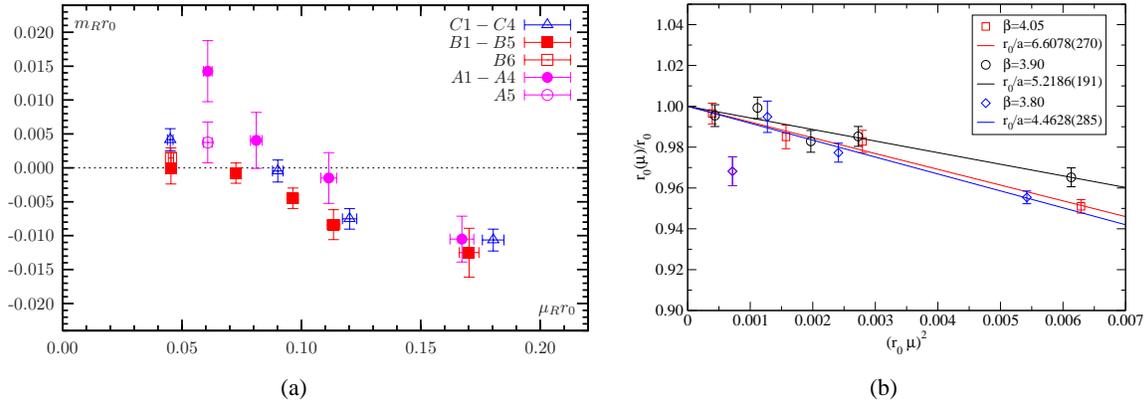}}
\hspace{0.37cm} 
\subfigure[]{\includegraphics[width=0.45\textwidth]{Figures/r0_vs_mu.eps}}
\vspace{-0.3cm}
\caption[]{(a) $m_{\rm R}r_0 = Z_A Z_P^{-1} m_{\rm PCAC} r_0$ vs.\
         $\mu_{\rm R}r_0$ and (b) $r_0(\mu)/r_0$ vs.\ $(\mu r_0)^2$,
         with $r_0 = \lim_{\mu \to 0} \; r_0(\mu)$ (see text). In both
         plots data for $\beta=4.05$, 3.9, 3.8 and $L \simeq 2.1, 2.1,
         2.4$~fm (respectively) are shown.}
\label{FIG:MTr0}
\end{center}
\end{figure}

Our practical implementation of maximal twist is illustrated by
Fig.~\ref{FIG:MTr0}a, where we show $m_{\rm R}r_0 \propto m_{\rm
PCAC}$ vs. $\mu_{\rm R}r_0 \propto \mu$ for $\beta=4.05$, $3.9$ and
$3.8$ (use of renormalized quantities eases the comparison).  Details
on $r_0/a$ and $Z_P$ (actually $Z_P(\overline{\rm MS},2~{\rm GeV})$)
are given below.

For $\beta=4.05$ and $3.9$ we could fulfill our criterion for maximal
twist at $\mu_{\rm R \; LOW} \simeq 0.047 r_0^{-1}$ with good
statistical precision: as shown in Fig.~\ref{FIG:MTr0}a, $m_{\rm
R}/\mu_{\rm R \; LOW} = Z_A m_{\rm PCAC}/\mu_{\rm LOW}$ is consistent
with zero within the statistical error, that we call
$\epsilon/\mu_{\rm LOW}$.  As a rule of thumb, we demand
$\epsilon/\mu_{\rm LOW}$ to be such that, numerically, $a\Lambda_{\rm
QCD} \epsilon/\mu_{\rm LOW} \lesssim 0.01$.  Considering the form of
the $d=5$ term in the Symanzik effective Lagrangian one finds in fact
that $a\Lambda_{\rm QCD} \epsilon/\mu$ is the expected order of
magnitude of the unwanted (relative) cutoff effects that may
contaminate physical observables if, due to numerical error, $m_{\rm
PCAC}(\beta,\mu)$ takes the value $\epsilon Z_A^{-1}$, rather than
zero.  Having no a priori control on the coefficients of this order of
magnitude estimate, we can only learn from the numerical experience of
a scaling test (involving our statistically most precise observables,
{\it i.e.}\ $f_{\rm PS}$ and $m_{\rm PS}$) whether our rule of thumb
yields a sufficiently accurate tuning to maximal twist.
\vspace{-0.3cm}

At $\beta=3.8$ our implementation of maximal twist was not as precise
as wished due to the long-range statistical fluctuations that, for
$\mu_R \lesssim 0.1r_0^{-1}$, were observed in our MC simulations (see
Ref.~\cite{CU-PROC} for information on autocorrelation times).  These
fluctuations, which increase in MC-time length and (weakly) in
amplitude as $\mu$ decreases, have a major impact on those
observables, such as the plaquette and $am_{\rm PCAC}$, that are not
continuous at the Singleton-Sharpe (1$^{\rm st}$ order) lattice phase
transition~\cite{AS-REP,SHARPE05}, rendering somewhat problematic the
estimate of their statistical errors.  This is manifest in
Fig.~\ref{FIG:MTr0}a from the two non-coinciding points obtained for
the lowest $\mu_{\rm R}$ at $\beta=3.8$, coming from two independent
simulations at slightly different $L$ (ensembles A1 and A5 in tab.~1
of Ref.~\cite{CU-PROC}).  In this case, the quoted (possibly
underestimated) statistical errors are larger than required by our
rule of thumb.

\subsection{Evaluation of $r_0$}
\label{r0eval}

In our scaling analyses we employ the Sommer scale
$r_0$~\cite{SOMMER93} to eliminate the lattice spacing $a$.  This is
meant only as an intermediate step.  In the end $r_0$ will be
eliminated in favour of $f_\pi$ ({\em i.e.}  $f_{\rm PS}$ at the
physical point) to get the scale for all dimensionful quantities. The
value of $r_0$ for the various ensembles was obtained with a better
than 0.5\% accuracy, starting from Wilson loops made out of
HYP-smeared temporal links~\cite{MDMetal} and APE-smeared spatial
links.  Employing several interpolating operators for static
quark-antiquark ($Q\bar{Q}$) states, corresponding to different
spatial smearings, leads to a matrix of correlators, from which the
(lowest) levels of the potential $V_{Q\bar{Q}}(r)$ are estimated by
solving a generalized eigenvalue problem.  A fit to the $r$-dependence
of the ground state gives $r_0(\mu)/a$.  For each $\beta$, the latter
shows a rather mild $\mu$-dependence, which is well described by a
first order polynomial in $\mu^2$.  The following values, extrapolated
to the chiral limit and denoted simply as $r_0/a$, will be used in
this work (those at $\beta=4.05$ and $\beta=3.8$ must be viewed as
preliminary): \\
\vspace{-0.7cm}
\begin{center}
$r_0/a|_{\beta=4.05} = 6.61(3) \, , \quad$ 
$r_0/a|_{\beta=3.9}  = 5.22(2) \, , \quad$ 
$r_0/a|_{\beta=3.8}  = 4.46(3) \, .$ \\
\end{center}
\vspace{-0.2cm}
Note that a fit by a second order polynomial in $\mu$ gives compatible
results.  An overview of the results for $r_0(\mu)$ for all $\beta$'s
is given in Fig.~\ref{FIG:MTr0}b, where the axes are normalized in
terms of the appropriate chirally extrapolated $r_0$. For $\beta=3.8$
the point at the lowest $\mu$-value (ensemble A1) is preliminary and
carries a still poorly estimated statistical error. It was hence not
used in the extrapolation to the chiral limit.  More details will be
given in a forthcoming publication.

\section{Charged pion sector: general remarks}
\label{generalities}

In the charged pseudoscalar (PS) meson sector our raw results for
$f_{\rm PS}$ and $m_{\rm PS}^2$ from simulations at $\beta=4.05$ and
$3.9$ with $L>2$~fm, once expressed in units of $r_0$, exhibit an
excellent scaling behaviour, see Fig.~4 of~Ref.~\cite{CU-PROC}. Also
the corresponding results at $\beta=3.8$, in spite of the
uncertainties on the implementation of maximal twist and the estimate
of statistical errors, appear consistent with a very good scaling
behaviour in the mass range $m_{\rm PS}=(350 \div 600)$~MeV.  These
findings are in agreement with the expectation that cutoff effects are
small in the absence of O($a$) artifacts. In particular in the charged
PS-meson sector it is known~\cite{ShWu2,FMPR} that, for small quark
masses $\mu_{\rm R}$, $m_{\rm PS}^2$ differs from its continuum
counterpart only by terms of O($a^2\mu_{\rm R}$) and O($a^4$), while
$f_{\rm PS}$ has discretization errors of O($a^2$).  This property
holds for any volume $L^3$ (sufficiently large to make pions much
lighter than other states) and is not affected by the lattice artifact
on the neutral PS-meson mass.\,\footnote{The relation of a possibly
large and negative lattice artifact on $m_{\pi^0}^2$ to
metastabilities in MC simulations has been discussed in
Refs.~\cite{ETM1,AS-REP}. Such metastabilities are not observed in the
simulations we consider here~\cite{CU-PROC}.}  In other words, to
order $a^2$ the cutoff effects on $m_{\rm PS}^2$ and $f_{\rm PS}$ are
like in a chirally invariant lattice formulation.  The use of
continuum chiral formulae to describe the volume and quark mass
dependences of our data in the charged PS-meson sector is thus well
justified.

\subsection{Estimates of $Z_P$ and of the renormalized quark mass}
\label{ZPestim}

In the study of the scaling of the renormalized quark mass $\mu_R$ one
needs the renormalization constant $Z_P(\beta;aq)$ at a common scale
$q$ for all the considered values of $\beta$. This renormalization
constant, as well as the scale-independent one $Z_A(\beta)$, was also
employed in Fig.~\ref{FIG:MTr0}a.\,\footnote{ For the renormalization
constants we keep the names they are given in the literature for the
standard (untwisted) Wilson quark lattice formulation, as obviously
their values do not change with respect to the untwisted Wilson case.}
Preliminary O($a$) improved estimates of the renormalization constants
of quark bilinear operators are reported in Ref.~\cite{PD-PROC}.
There the results for $Z_A$ are rather precise (at the 1.5\% level),
while the quoted uncertainties on $Z_P$ are still substantially
larger.

In view of this situation, in our scaling analysis of $\mu_R$ we do
not use the values of $Z_P$ quoted in Ref.~\cite{PD-PROC}. We rather
employ the scale- and scheme-independent ratios
$Z_P(\beta;aq)/Z_P(\beta_{\rm ref};a_{\rm ref}q)$, which we extract
with a statistical accuracy of $\sim 1\%$ from the relation (exact up
to O($a^2$) terms)
\begin{equation}
Z_P(\beta;aq) \, / \,Z_P(\beta_{\rm ref};a_{\rm ref}q) \;\, = \;\,
\mu(\beta; m_{\rm PS}r_0=1; L/r_0 \simeq 5) \,/ \,
\mu(\beta_{\rm ref}; m_{\rm PS}r_0=1;L/r_0 \simeq 5) \, ,
\label{match_pimass}
\end{equation}
in order to compare the values of $\mu_R(\beta;aq)Z_P(\beta_{\rm
ref};a_{\rm ref}q)$ at different values of $\beta$ and of the PS-meson
mass.  In eq.~(\ref{match_pimass}), $a\mu(\beta; m_{\rm PS}r_0=1;
L/r_0 \simeq 5)$ is the value of $a\mu$ for which, at a given $\beta$
and $L/r_0 \simeq 5$, one finds $m_{\rm PS}r_0=1$.\,\footnote{This
value of $m_{\rm PS}r_0$ was chosen so as to lie in the region where
errors from interpolation to the reference mass and cutoff effects are
smallest.}  The scaling is obviously not affected by the extra overall
factor $Z_P(\beta_{\rm ref};a_{\rm ref}q)$, which is removed in the
end (in order to give an idea of the values of the renormalized quark
masses).  In the following we choose $\beta_{\rm ref}=3.9$, where we
most reliably know (see Ref.~\cite{PD-PROC}) $Z_P(\beta_{\rm
ref};a_{\rm ref}q)$, we evaluate for $\beta=4.05$ and $\beta=3.8$ the
$Z_P$-ratios from eq.~(\ref{match_pimass}) and we finally obtain
estimates of $\mu_R(\beta;aq)$ at $q=2~{\rm GeV}$ in the
$\overline{\rm MS}$ scheme for all $\beta$'s. An important drawback of
this method is the fact that cutoff effects stemming from data
obtained at different $\beta$-values mix up in the quark mass
renormalization, which in general may fake the genuine $a$-dependence
and cast doubts on the reliability of any continuum extrapolation.
Nevertheless, since our data for $\beta=4.05$ and $3.9$ show no
statistically significant cutoff effects in the relation of $f_{\rm
PS}$ to $m_{\rm PS}^2$ and in the values of
$\mu_R(\beta;aq)Z_P(\beta_{\rm ref};a_{\rm ref}q)$, we can obtain an
estimate of the continuum limit of $\mu_R$ (in units of $r_0$) by
taking an average of the results at these two $\beta$-values.  A
similar remark holds for the determination of $\widehat{B}_0$ and the
chiral condensate in sect.~\ref{physres}.  As discussed below, for all
observables we associate to our continuum limit estimates a
conservative systematic error obtained by comparing them to the
corresponding results from data at $\beta=3.8$.

\subsection{About taking the continuum, thermodynamical and chiral limits}
\label{strategy}

It is well known that matching simulation data obtained at finite $L$
and for $m_{\rm PS} \geq 300$~MeV to the physical pion point requires
a delicate analysis. The impact of residual O($a^2$) effects (even if
small) on such an analysis might be enhanced if the continuum limit is
performed as the last step. It is therefore advisable to perform first
a continuum limit extrapolation at different (suitably chosen) fixed
physical conditions and then use Chiral Perturbation Theory ($\chi$PT)
to correct for finite size effects and reach the physical pion point.
This is the strategy we follow in sect.~\ref{contfit}, based on the
results of the scaling test presented in sect.~\ref{scalingcont}.  As
far as one is concerned with data, such as those at $\beta=4.05$ and
$3.9$, where no significant cutoff effects are observed, a
conceptually equivalent approach is that of performing a combined
analysis of all data by means of continuum $\chi$PT formulae.  The
outcome of this approach, which is detailed in Ref.~\cite{CU-PROC}, is
summarised in sect.~\ref{combfit}.  For comparison we also discuss in
sect.~\ref{singfit} the results of fits to continuum $\chi$PT formulae
where the data corresponding to different lattice spacings are treated
separately.

\section{Charged pion sector: scaling test}

Here we analyse the scaling behaviour of the charged PS-meson decay
constant and the renormalized quark mass as $a \to 0$ at fixed values
of $m_{\rm PS}r_0$ and $L/r_0$. The renormalization and scaling
conditions are as follows:
(i) $r_0^2 F_{Q\bar{Q}}(r_0) = 1.65$ (with 
   $F_{Q\bar{Q}}$ the static interquark force), which
  allows to trade $g_0^2=6/\beta$ with $r_0/a$ ; 
(ii) 
 $m_{\rm PS} r_0 = {\rm constant} \in \{0.7, 0.8, 0.9, 1.0, 1.1, 1.25\}$, 
  which eliminates  $\mu$ in favour of $m_{\rm PS}$ ;
(iii) fixed spatial volume 
    $L_{\rm ref}^3 \simeq (2.2~{\rm fm})^3$,
    which corresponds to $L/r_0 \simeq 5$. 

\subsection{Scaling and preliminary continuum limit estimates}
\label{scalingcont}

The typical relative statistical errors are (conservatively) estimated
to be about 0.7\% for $m_{\rm PS}r_0$, 1.0\% for $f_{\rm PS}r_0$ and
1.5\% for $\mu_{\rm R}r_0$.  In the latter case, besides the error on
$r_0/a$, only the uncertainty (typically about 1.2\%) on the
$Z_P$-ratios (see eq.~(\ref{match_pimass})), is taken into account and
hence shown in Fig.~\ref{FIG:mu_R}. The error on $Z_P(\beta_{\rm
ref};a_{\rm ref}q)$ is omitted here, as it plays no role, while in
sect.~\ref{physres} it is taken into account, as it matters for the
final values of $\widehat{B}_0$, the chiral condensate and $m_{ud}$.

\begin{figure}[!t]
\begin{center}
\subfigure[]{\includegraphics[scale=0.29,angle=0]{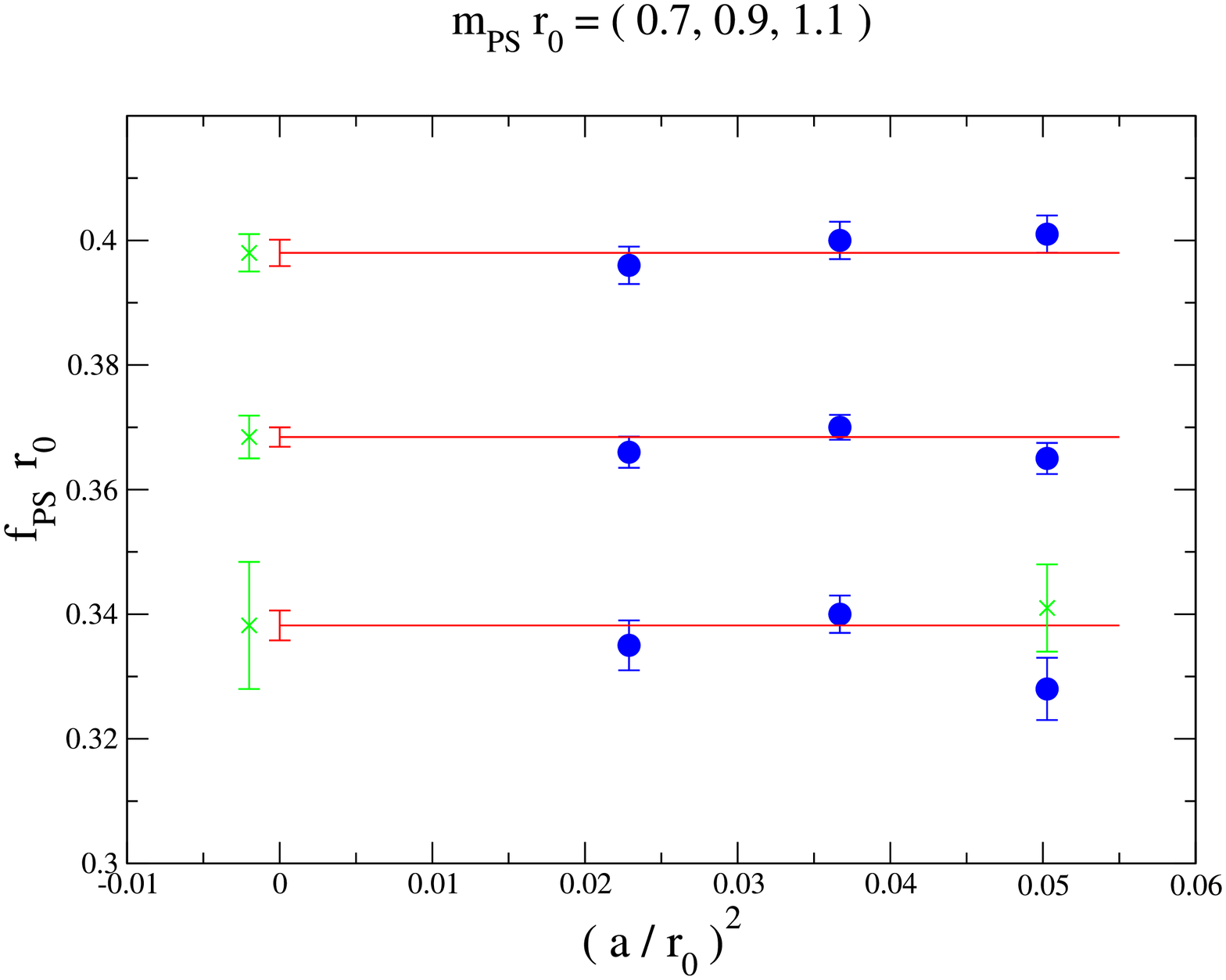}}
\subfigure[]{\includegraphics[scale=0.29,angle=0]{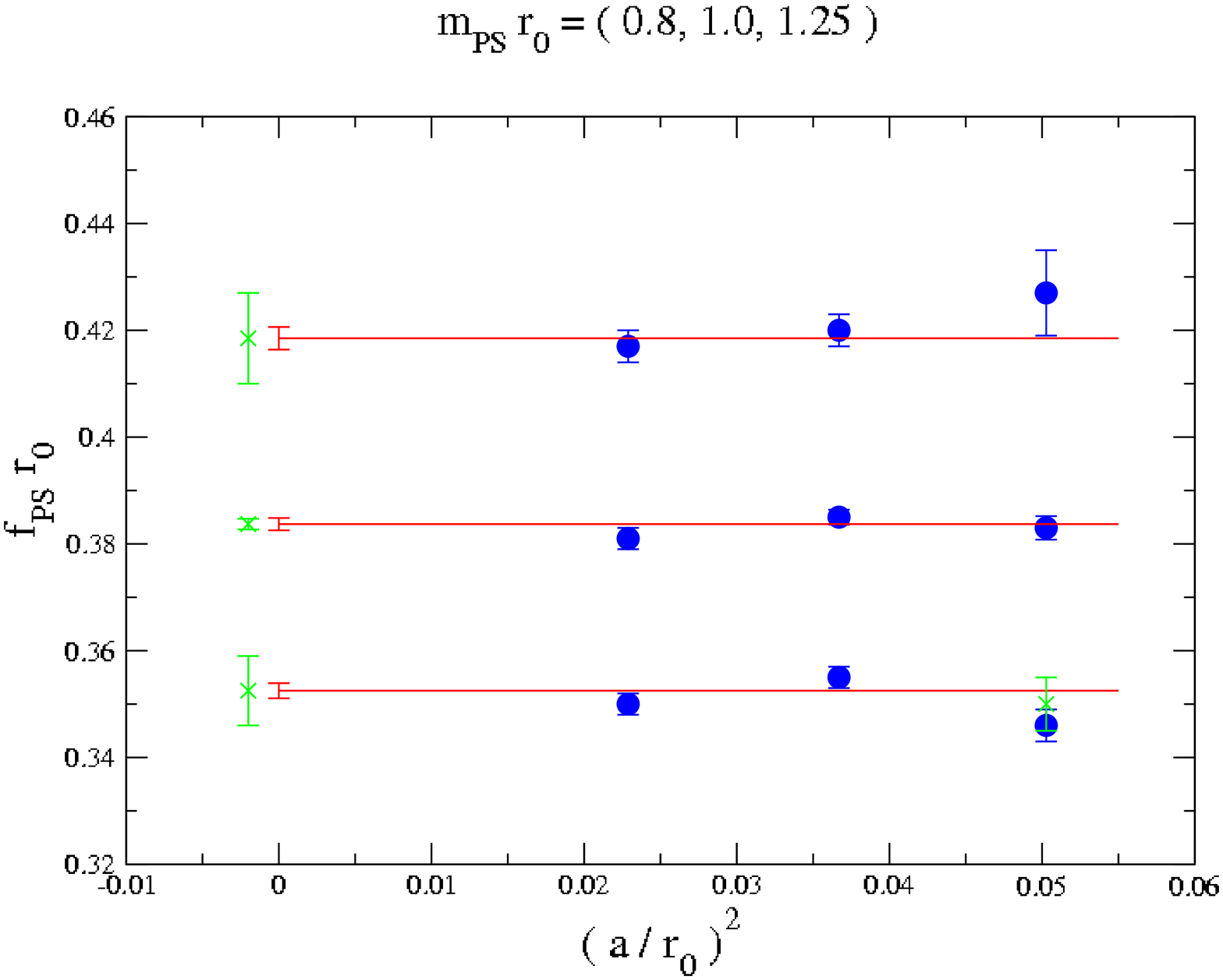}}
\vspace*{-0.3cm}
\caption[]{Scaling plots of $f_{\rm PS}r_0$ vs. $(a/r_0)^2$ at fixed
  values of $m_{\rm PS}r_0$ (increasing from bottom to top).  See text
  for a detailed explanation of the symbols.}
\label{FIG:f_ps}
\end{center}
\end{figure}

\begin{figure}[!t]
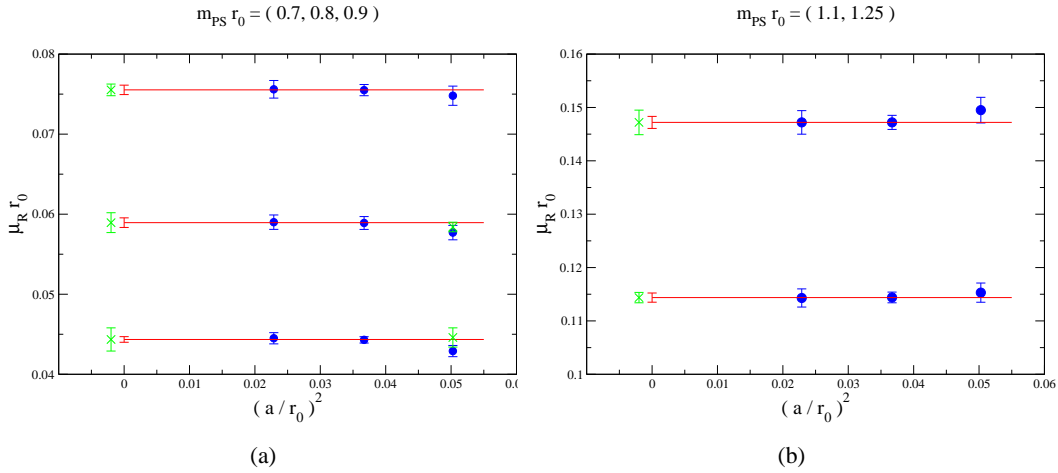

\begin{center}
\subfigure[]{\includegraphics[scale=0.29,angle=0]{Figures/muR_plotA.eps}}
\subfigure[]{\includegraphics[scale=0.29,angle=0]{Figures/muR_plotB.eps}}
\vspace*{-0.3cm}
\caption[]{Scaling plot of $\mu_{\rm R}r_0$ vs. $(a/r_0)^2$ at fixed
  values of $m_{\rm PS}r_0$ (increasing from bottom to top).  See text
  for a detailed explanation of the symbols.}
\label{FIG:mu_R}
\end{center}
\end{figure}

The condition (i) is immediately fulfilled once all quantities are
expressed in units of $r_0$. As for the other two conditions, we first
implement the condition (iii) by ``moving'' via resummed $\chi$PT
formulae (see Ref.~\cite{Colangelo:2005gd}) the data for $f_{\rm PS}$
and $m_{\rm PS}$ from the simulation volume $L^3$ to the reference one
$L_{\rm ref}^3$.  As simulations were done with $L \in (2.0, 2.4)$~fm,
the numerical change in the data is very tiny (never larger than
0.7\%) and thus statistically almost irrelevant.  Then, in order to
match the reference values of $m_{\rm PS}r_0$ given in (ii), we
perform interpolations (in few cases also short extrapolations) of the
values of $f_{\rm PS}r_0$ and $\mu r_0$. For this purpose we try both
(low order) polynomial and $\chi$PT-inspired fits to the data. The
spread among the different fits with good $\chi^2$, whenever
statistically significant, is added linearly to the interpolation
error.  In this way one ends up with the blue filled circle data
points (and errorbars) in the Figs.~\ref{FIG:f_ps} and \ref{FIG:mu_R}.

For each value of $m_{\rm PS}r_0$ we obtain preliminary estimates of
the continuum limit values of $f_{\rm PS}r_0$ and $\mu_{\rm R}r_0$ by
fitting to a constant (red line in the figures) {\em only} the data
points from simulations at $\beta=4.05$ and $\beta=3.9$.  The results,
with only statistical error from the continuum extrapolation, are
shown in red in Figs.~\ref{FIG:f_ps} and \ref{FIG:mu_R}. In all cases
the difference between the result of the continuum extrapolation and
the central value (blue filled circle point) for $\beta=3.8$ is taken
as an estimate of the systematic error (indicated with a green cross
placed at a slightly negative value of $a$) on the continuum limit
result.  Since the latter is obtained by simply taking a weighted
average of the results at $\beta=4.05$ and $3.9$, the introduction of
such an error appears necessary. Moreover, the quality of the data in
the figures suggests that the way we estimate this systematic error is
rather conservative.  Finally, the green crosses (and errorbars)
appearing in Figs.~\ref{FIG:f_ps} and \ref{FIG:mu_R} at $(a/r_0)^2
\sim 0.05$ show, whenever the displacement is larger than one standard
deviation, where the actual data would move if one were correcting for
the leading effect of the numerical error (denoted by $\epsilon$ in
sect.~\ref{tuneMT}) in imposing $m_{\rm PCAC}=0$. We refer to the
results of such a correction (explained in sect.~\ref{rotMT}) as to
``data moved to maximal twist''.

\subsection{About ``moving data to maximal twist''}
\label{rotMT}

The procedure of ``moving data to maximal twist'' can be seen as a way
of testing whether statistical errors in the tuning to maximal twist
have any significant impact on the observables of interest. If one has
$Z_A m_{\rm PCAC}=\epsilon$ (rather than zero), the effective twist
angle in the Symanzik Lagrangian is $\alpha = \pi/2 - \theta$ with
$\tan \theta = m_R/\mu_R = \epsilon/\mu$. Treating $\epsilon$ as an
O($a^0$) quantity and neglecting O($a\theta$) and O($a^2$) effects,
the consequences of a deviation from maximal twist are in general
twofold. First, the effective renormalized quark mass becomes $M_R =
(\mu_R^2 + m_R^2)^{1/2} = \mu_R/cos\theta$.  Second, in all operator
matrix elements any operator formally non-invariant under
axial-$\tau^3$ transformations must be reinterpreted consistently with
the twist angle being $\alpha = \pi/2 - \theta$ (rather than
$\pi/2$). As a consequence, once the hadronic states of interest are
correctly identified, simple $\theta$-dependent formulae can be
derived that allow to extract properly the matrix elements of
operators non-invariant under axial-$\tau^3$ transformations.  For
instance, the (renormalized) operator $Z_V \bar\chi \gamma_\mu \tau^2
\chi$ (written in the basis where the Wilson term has its standard
form) coincides with $\cos\theta A_{\mu\;R}^1 + \sin\theta
V_{\mu\;R}^2$, where $A_{\mu\;R}^1$ ($V_{\mu\;R}^2$) is the physical
axial (vector) current. Consequently one finds $\;\; < \pi^1 | Z_V
\bar\chi \gamma_\mu \tau^2 \chi | \Omega >|_{m_R,\mu_R} \; = \;
\cos\theta < \pi^1 | A_{\mu\;R}^1 | \Omega >|_{M_R} + {\rm O}(a\theta,
a^2) \; $, which implies $\; m_{\rm PS}f_{\rm PS}|_{M_R} = < \pi^1 |
Z_V \bar\chi \gamma_0 \tau^2 \chi | \Omega >|_{m_R,\mu_R} /
\cos\theta + {\rm O}(a\theta, a^2)$.

From these arguments it follows that $m_{\rm PS}$-data need not be
moved, while the data for $\mu_{\rm R}$ and $f_{\rm PS}$ are ``moved
to maximal twist'', up to ${\rm O}(a\theta, a^2)$, by dividing them by
$\cos\theta$, where $\theta$ is obtained from the (interpolated)
actual values of $\epsilon$ and $\mu$ in our data sets.  It turns out
that the change implied by this correction is larger than one standard
deviation (of the uncorrected data) only for the two smallest
reference values of $m_{\rm PS}r_0$ at $\beta=3.8$. After such
correction all the data points at $\beta=3.8$ (including the two most
chiral ones) fit very nicely into the scaling picture suggested from
the corresponding data at $\beta=3.9$ and $\beta=4.05$.

However, any data where the above correction was statistically
relevant can be hardly used for a continuum extrapolation, because the
residual (uncorrected) O($a\theta$) lattice artifacts depend on the
details of the numerical errors at the different $\beta$-values and
hence need not to scale as $a \to 0$. Nevertheless, at a more
qualitative level, the procedure of ``moving data to maximal twist''
confirms that, even at $\beta=3.8$ ({\it i.e.}\ $a \simeq 0.1$~fm),
the lattice artifacts in the charged PS-meson sector appear to be
quite small, provided maximal twist is implemented precisely, e.g.\
according to the criteria of sect.~\ref{tuneMT}.  In fact such a
precise implementation is rather hard to achieve in the case of our
simulations at $\beta=3.8$, as we found $\tau_{\rm int}(am_{\rm PCAC})
= {\rm O}(100)$, see sect.~\ref{tuneMT} and Ref.~\cite{CU-PROC}.

\section{Charged pion sector: physical results}
\label{physres}

Here we discuss the description of our $m_{\rm PS}$- and $f_{\rm
PS}$-data by means of {\em continuum} $\chi$PT for $N_{\rm f}=2$ QCD
and give preliminary estimates of the low energy constants (LEC)
$\bar{l_3}$, $\bar{l_4}$, $\widehat{B}_0$ and $f_0$,\,\footnote{ We
use the convention $f_0 = \sqrt{2} F_0$, {\it i.e.} the normalization
$f_\pi = 130.7~{\rm MeV}$.}  as well as of the average light quark
mass, $m_{ud}$, and the chiral condensate.

We employ the following continuum NLO $\chi$PT
formulae~\cite{Gasser:1986vb,Colangelo:2005gd} to simultaneously
describe the dependence of $m_{\rm PS}$ and $f_{\rm PS}$ on the bare
quark mass ($\mu$) and on the finite spatial size ($L$):
\begin{equation}
  m_\mathrm{PS}^2(L) = 2B_0\mu \, K_m^2(L) \, \left[ 1 +
    \xi \ln ( 2B_0\mu/\Lambda_3^2 ) \right] \, , 
  \label{eq:chirfo1} 
\end{equation}
\begin{equation} 
  f_\mathrm{PS}(L) = f_0 \,K_f(L)  \, \left[ 1 -
    2 \xi \ln ( 2B_0\mu/\Lambda_4^2 ) \right]  \, ,
  \label{eq:chirfo2}
\end{equation}
\noindent where~~$\xi = 2B_0\mu/(4\pi f_0)^2$ and $K_{m,f}(L)$ account
for finite size (FS) effects.  The LEC $\bar{l}_{3,4}$ are related to
the parameters $\Lambda_{3,4}$ introduced in
eqs.~(\ref{eq:chirfo1})--(\ref{eq:chirfo2}) through $\bar{l}_{3,4}
\equiv \log(\Lambda_{3,4}^2/m_{\pi^{\pm}}^2)$.  The expressions for
$K_{m,f}(L)$ at NLO (denoted as GL) read~\cite{Gasser:1986vb}:
$K_m^{\rm GL}(L) = 1 + \frac{1}{2}\xi \tilde{g}_1( \lambda )$ and
$K_f^{\rm GL}(L) = 1 - 2 \xi \tilde{g}_1( \lambda )$, where $\lambda =
\sqrt{2B_0\mu L^2}\,$ and $\tilde{g}_1( \lambda )$ is a known
function. A convenient way to include higher order $\chi$PT terms in
the description of FS effects on $m_{\rm PS}$ and $f_{\rm PS}$ is
provided by the formulae of Ref.~\cite{Colangelo:2005gd} (denoted by
CDH).  In the following we always use the CDH expressions for
$K_{m,f}(L)$, which turn out~\cite{CU-PROC} to describe well our data
at different volumes (and for $m_{\rm PS}$ better than GL formulae).

In all the analyses below physical units are introduced as follows:
the experimental values, $f_{\pi}=130.7$~MeV and $m_{\pi^0}=
135.0$~MeV,\,\footnote{The $\pi^0$-mass input is chosen in view of the
absence of electromagnetic effects in our lattice QCD simulations.}
are exploited to determine first the $\mu$-value, $\mu_\pi$,
corresponding to the ``physical point'' through
$m_\mathrm{PS}/f_\mathrm{PS}|_{\mu_\pi} = m_{\pi^0}/f_{\pi}$, and then
(depending on the analysis) the value of $r_0f_{\pi}$, or $af_\pi$.
The latter allows to express all quantities in units of $f_\pi$.

\subsection{Chiral fits and continuum estimates of LEC}

As anticipated in sect.~\ref{strategy}, given the good scaling
behaviour of our data, we present {\em preliminary} continuum
estimates of LEC and $m_{ud}$ that stem from two analyses -- see
sects.~\ref{contfit} and \ref{combfit}.  In sect.~\ref{singfit} we
give the corresponding results that are obtained at two fixed lattice
spacings, $a(\beta=4.05)$ and $a(\beta=3.9)$. The latter results will
serve mainly for estimating (in the way detailed below) the systematic
error due to residual O($a^2$) cutoff effects.

Concerning other systematic errors, we note: (i) residual
uncertainties in the CDH-formulae for FS effects are small, compared
to other systematic errors -- see the discussion in
Ref.~\cite{CU-PROC}; (ii) the possible impact of NNLO corrections to
the formulae~(\ref{eq:chirfo1})--(\ref{eq:chirfo2}) for the quark mass
dependence is minimized by taking out of the analyses the points with
highest values of $m_{\rm PS}$ (we checked that fits are stable if we
leave out data with $m_{\rm PS} > 500$~MeV -- more details in a
forthcoming publication).

\subsubsection{$\chi$PT fits in the continuum}
\label{contfit}

Following the strategy presented in sect.~\ref{strategy} one can
estimate the continuum limit values of $f_\mathrm{PS}r_0$ and $\mu_R
r_0$ at fixed values of $m_\mathrm{PS}r_0$ and $L_{\rm ref}/r_0$, as
reported in sect.~\ref{scalingcont}.  Considering $f_{\rm PS}r_0$ and
$m_\mathrm{PS}r_0$ as functions of $\mu_R r_0$ allows for direct use
of the $\chi$PT formulae~(\ref{eq:chirfo1})--(\ref{eq:chirfo2}) to
bring our data from $L_{\rm ref}=2.2$~fm to infinite volume and
parameterise their quark mass dependence. Leaving out the point
corresponding to $m_{\rm PS}r_0 =1.25$, the data are well described by
our fit ansatz and we obtain the following values for the fit
parameters:
\begin{eqnarray}
    2\widehat{B}_0r_0\quad  = \quad 12.0(3)(7)\, , 
    \quad \quad &\bar{l}_3\quad  = \quad 3.67(12)(35)\, , 
    \nonumber \\
    f_0r_0\quad =\quad 0.266(3)(10)\, , 
    \quad \quad &\bar{l}_4\quad =\quad 4.69(4)(11)\, ,
    \label{eq:bestfitcont1}
\end{eqnarray}
where the renormalized quantity $\widehat{B}_0=Z_P B_0$ is given in
the $\overline{\rm MS}$-scheme at the scale $q=2$~GeV (see
sect.~\ref{ZPestim}). The $\chi^2/{\rm dof}$ of the fit is 0.28. Using
the experimental input at the physical pion point we find
$r_0=0.433(5)(16)$~fm.  Inserting the values of $r_0/a$ (see
sect.~\ref{r0eval}), one gets the estimates\\
\vspace*{-0.4cm} \\
\hspace*{0.1cm} $a|_{\beta=4.05}=0.0655(8)(24)$~fm,
$\quad a|_{\beta=3.9}=0.0830(10)(31)$~fm,  
$\quad a|_{\beta=3.8}=0.0970(13)(37)$~fm. \\
\vspace*{-0.4cm}

In the above results, except for the case of $2\widehat{B}_0r_0$, the
second error comes entirely from the systematic uncertainty due to
residual O($a^2$) cutoff effects: for each quantity this error is
conservatively taken as the maximum of (i) the uncertainty resulting
from the propagation through the chiral analysis of the systematic
error associated to the continuum estimates as derived in
sect.~\ref{scalingcont} and (ii) the spread of the results obtained
separately at $\beta=3.9$ and $\beta=4.05$ (by fitting to the same
$\chi$PT ansatz as here -- see sect.~\ref{singfit}). In practice the
maximum is usually given by the former of these two systematic error
estimates.  Note however that for $\widehat{B}_0$, $m_{ud}$ and the
chiral condensate, an additional systematic uncertainty coming from
$Z_P$ (as quoted in Ref.~\cite{PD-PROC}) is added in quadrature.

As a check, we also study the decay constant $f_\mathrm{PS}$ as a
function of $m_\mathrm{PS}$, {\it i.e.}  with no reference to
$\mu_{\rm R}$. The appropriate NLO $\chi$PT fit ansatz is obtained
(ignoring NNLO corrections) by replacing $2B_0\mu$ with
$m_\mathrm{PS}^2$ in eq.~(\ref{eq:chirfo2}).  The resulting best fit
parameters are $f_0 r_0 = 0.268(3)(12)$ and $\bar{l}_{4} =
4.82(4)(14)$.  This is consistent with the values in
eq.~(\ref{eq:bestfitcont1}) and yields for the Sommer scale the
estimate $r_0=0.435(4)(15)~{\rm fm}$. The second error again comes
from (the propagation of) the systematic uncertainty in the
``continuum extrapolation'' of sect.~\ref{scalingcont}.

\subsubsection{$\chi$PT analysis combining $\beta=3.9$ and $4.05$}
\label{combfit}

As argued in sect.~\ref{strategy}, given the absence of statistically
relevant cutoff effects, one can combine the data (in lattice units)
for $m_{\rm PS}$ and $f_{\rm PS}$ coming from our simulations at
$\beta=3.9$ and $\beta=4.05$ and perform a global combined fit based
again on eqs.~(\ref{eq:chirfo1})--(\ref{eq:chirfo2}). Such a combined
fit has six free parameters which can be taken
as~~$(aB_0)|_{\beta=3.9}$, \ $(aB_0)|_{\beta=4.05}$,
$(af_0)|_{\beta=3.9}$, $(af_0)|_{\beta=4.05}$, $\Lambda_3/f_0$ and
$\Lambda_4/f_0$. Details and plots concerning this analysis can be
found in Ref.~\cite{CU-PROC}.  The outcome is summarised in
table~\ref{TAB_1}.\,\footnote{From this analysis, without using $r_0$,
we find $a|_{\beta=3.9}/a|_{\beta=4.05} = 1.28(1)$, which is quite
close to $(r_0/a)|_{\beta=4.05}/(r_0/a)|_{\beta=3.9}=1.27(1)$.
Moreover, the comparison of the ratios $f_0/B_0|_{\beta=3.9}$ and
$f_0/B_0|_{\beta=4.05}$ provides a rather precise estimate of the
ratio
$Z_P(\beta=3.9;a_{\beta=3.9}q)/Z_P(\beta=4.05;a_{\beta=4.05}q)$.} As
for the estimates of $a$, the second error we quote is an estimate
(see sect.~\ref{contfit}) of the systematic uncertainty due to
residual cutoff effects.  We note that, leaving out the cases for
which $m_{\rm PS} > 500$~MeV, nine ensembles of gauge configurations
enter this analysis: five at $\beta=3.9$ ($B_1$ to $B_4$ and $B_6$)
and four at $\beta=4.05$ ($C_1$ to $C_3$ and $C_5$).
\begin{table}[!t]
  \centering
  \begin{tabular*}{0.9\textwidth}{@{\extracolsep{\fill}}c|c|c|cc}
    \hline\hline
    & single-$\beta$ fit      & single-$\beta$ fit      & \multicolumn{2}{c}{combined fit -- see sect.~\ref{combfit}}\\
    $\Bigl.\Bigr.\beta$   & $3.9$        & $4.05$       & $3.9$            & $4.05$          \\
    \hline \hline
    $2aB_0$               & $4.85(4)$    & $3.87(6)$    & $4.87(4)$        & $3.76(3)$       \\
    $a f_0$               & $0.0526(4)$  & $0.0404(7)$  & $0.0527(4)$      & $0.0411(4)$     \\ 
    $\Lambda_3/f_0$       & $6.36(26)$   & $7.20(48)$   & \multicolumn{2}{c}{$6.41(26)$ }    \\ 
    $\Lambda_4/f_0$       & $11.59(19)$  & $11.81(31)$  & \multicolumn{2}{c}{$11.51(21)$}    \\ 
    $\chi^2/{\rm dof}$    & $7.8/6$      & $2.2/4$      & \multicolumn{2}{c}{$12.0/12$}      \\
    \hline
    $a \mu_\pi$           & $0.00072(2)$ & $0.00054(2)$ & $0.00072(1)$     & $0.00057(1)$    \\
    $a~[{\rm fm}]$        & $0.0854(6)$  & $0.0656(10)$ & $0.0855(5)(31)$  & $0.0667(5)(24)$ \\
    \hline \hline
  \end{tabular*}
  \caption{Results from $\chi$PT fits at single $\beta$-values and
    from the global combined fit of
     sect.~4.1.2.
  }
  \label{TAB_1}
\end{table}

\subsubsection{Independent $\chi$PT analyses at $\beta=3.9$ and $4.05$ and comparison}
\label{singfit}

\begin{figure*}[!t]
  \begin{center}
    \subfigure[]{\includegraphics[width=.47\textwidth]{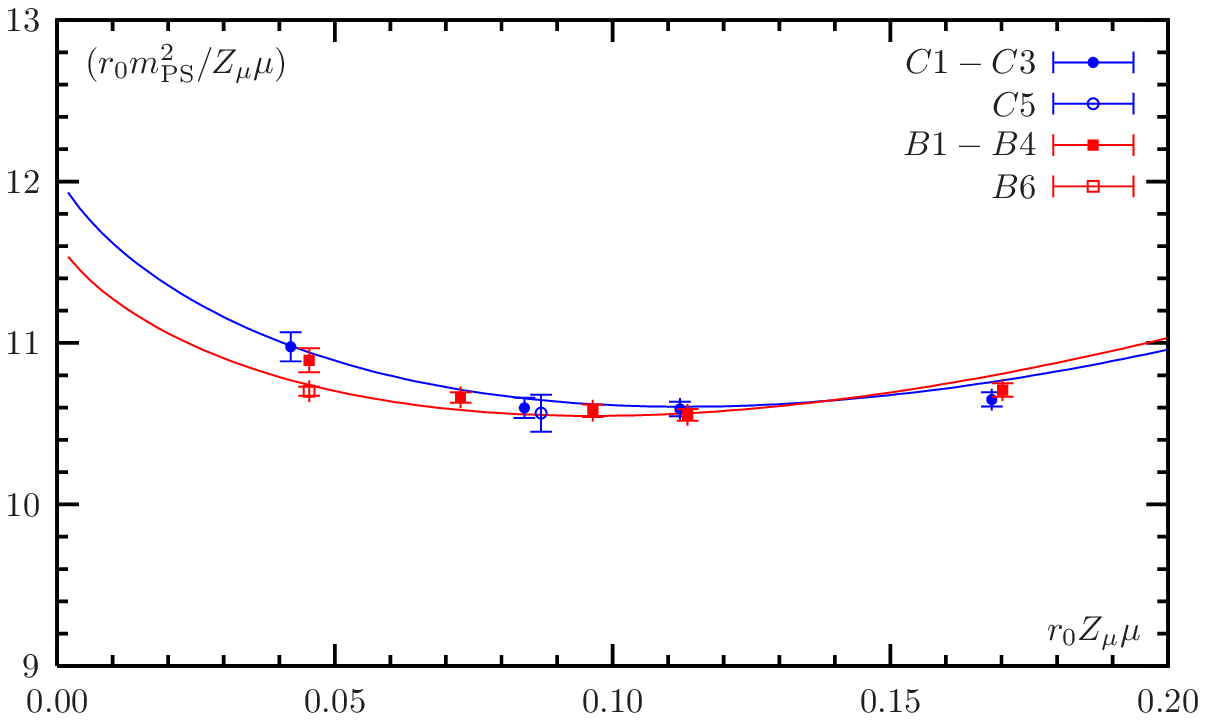}}
    \hspace*{0.5cm}
    \subfigure[]{\includegraphics[width=.47\textwidth]{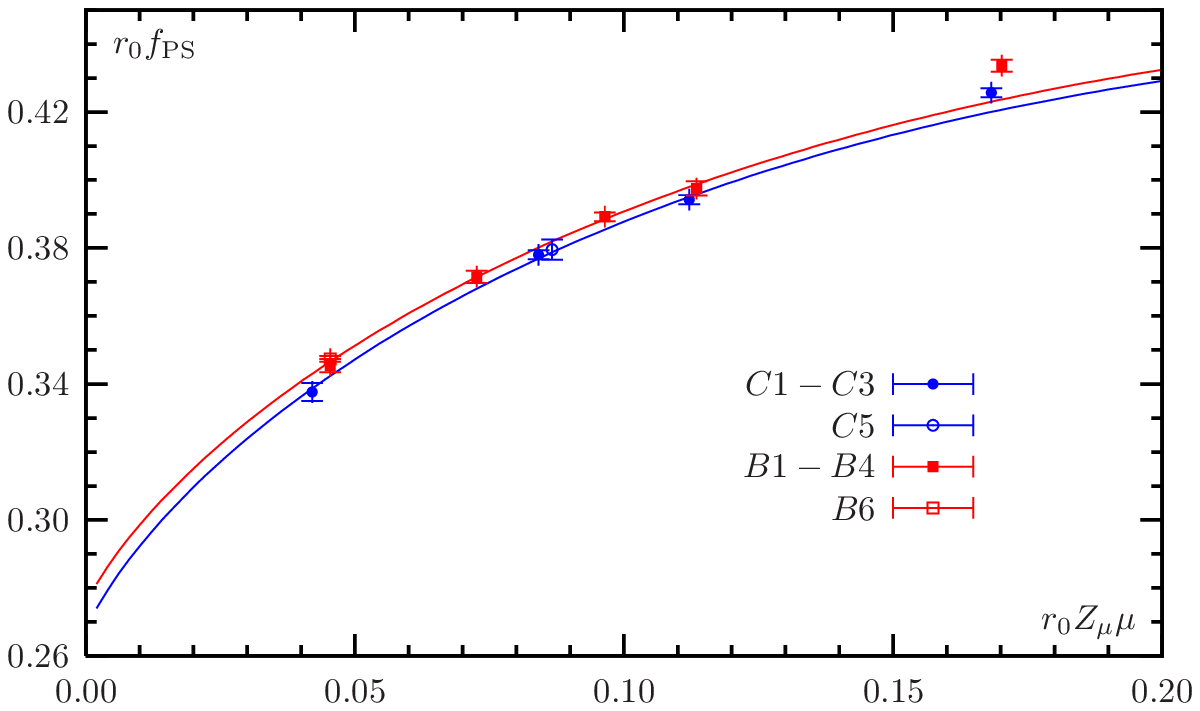}}
    \vspace*{-0.3cm}
    \caption{
      Independent $\chi$PT fits at $\beta=4.05$ and $\beta=3.9$: 
      FS-corrected data \`a la CDH for (a) $r_0m_{\rm PS}^2/\mu_{\rm R}$ 
      and (b) $r_0f_{\rm PS}$ vs.\ $\mu_{\rm R}r_0$ are shown, with the corresponding 
      best fit curves to eqs.~(4.1)--(4.2).
      The points at $\mu_{\rm R}r_0 \sim 0.17$ were not included in the fit.
    }
    \label{chifits}
  \end{center}
\end{figure*}

These analyses, always based on
eqs.~(\ref{eq:chirfo1})--(\ref{eq:chirfo2}), are closely analogous to
that presented in Ref.~\cite{ETM1}.  The free parameters can be taken
to be $aB_0$, $af_0$, $\Lambda_3/f_0$ and $\Lambda_4/f_0$, the best
fit values of which are given in table~\ref{TAB_1} and illustrated in
Fig.~\ref{chifits}.
For the case of $\beta=3.9$ the only difference with respect to
Ref.~\cite{ETM1} is the increased statistics and in particular the
presence of the ensemble $B_6$.  For both $\beta=4.05$ and $\beta=3.9$
the same ensembles are considered as in the case of the global
combined fit of sect.~\ref{combfit}.  It should be noted that in the
analysis of sect.~\ref{contfit} the ensembles $C_5$ and $B_6$ were not
employed, in order to avoid to heavily rely on the CDH-formulae in the
``continuum extrapolations'' discussed in sect.~\ref{scalingcont}. On
the other hand using these two ensembles, in particular $B_6$ (which
corresponds to $L \sim 2.7$~fm) increases the statistical information.

In table~\ref{TAB_2} results from the independent analyses at
$\beta=3.9$ and $4.05$, the combined fit of sect.~\ref{combfit} and
the fit in the continuum of sect.~\ref{contfit} are compared (the
second error, whenever quoted, is the systematic one previously
discussed).  The agreement between the last two columns of
table~\ref{TAB_2} is good, even within statistical errors only. We
recall that the ensembles $B_6$ and $C_5$ do not contribute to the
results in the last column. Were these ensembles not used also in the
combined fit of sect.~\ref{combfit}, the results in the third column
of table~\ref{TAB_2} would get, as expected, even closer to those in
the last column: $\bar{l}_3 = 3.66(9)$, $\bar{l}_4 = 4.64(5)$, $2r_0
\widehat{B}_0=11.9(3)$, $f_0 r_0 = 0.270(3)$, $r_0 = 0.439(5)$~fm. For
reviews on determinations of the LEC we refer to
Refs.~\cite{Leutwyler:2007ae,SN-PROC}.
\begin{table}[!t]
  \centering
  \begin{tabular*}{1.0\textwidth}{@{\extracolsep{\fill}}c|c|c|cc|c}
    \hline\hline
    & single-fit at $\beta=3.9$  & single-fit at $\beta=4.05$ & \multicolumn{2}{c|}{combined fit} & fit in the continuum   \\
    \hline \hline
    $\bar{l}_3$           & $3.41(9)$          & $3.66(14)$         & \multicolumn{2}{c|}{$3.44(8)(35)$}  & $3.67(12)(35)$ \\ 
    $\bar{l}_4$           & $4.62(4)$          & $4.66(7)$          & \multicolumn{2}{c|}{$4.61(4)(11)$}  & $4.69(4)(11)$  \\
    \hline
    $2r_0 \widehat{B}_0$  & $11.6(3)(5)$       & $11.9(3)(5) \div 12.1(3)(5)$   & \multicolumn{2}{c|}{$11.6(3)(7)$}   & $12.0(3)(7)$   \\ 
    $f_0 r_0$             & $0.275(2)$         & $0.267(5)$         & \multicolumn{2}{c|}{$0.273(3)(10)$} & $0.266(3)(10)$ \\
    \hline
    $r_0~[{\rm fm}]$      & $0.446(4)$         & $0.434(7)$         & \multicolumn{2}{c|}{$0.444(4)(16)$} & $0.433(5)(16)$ \\
    \hline \hline
  \end{tabular*}
  \caption{Estimates of LEC and $r_0$ from the analyses of
    sects.~4.1.3 (1$^{\rm st}$ and 2$^{\rm nd}$ column), 4.1.2
    (3$^{\rm rd}$ column) and 4.1.1 (last column).  The values of $2r_0
    \widehat{B}_0$ are obtained using $Z_P$ at $\beta=3.9$~\cite{PD-PROC}
    and $Z_P$-ratios from the analyses of sects.~4.1.2 and 4.1.1: at
    $\beta=4.05$ we quote the two (similar) results obtained
    in this way.}
  \label{TAB_2}
\end{table}

\subsection{Continuum estimates of $m_{ud}$ and the chiral condensate}
\label{m_chi}

Our preliminary results in the $\overline{\rm MS}$-scheme at scale
$q=2$~GeV are obtained using the RI-MOM estimate~\cite{PD-PROC} of
$Z_P$ at $\beta=3.9$ (converted to $\overline{\rm MS}$ by using NNNLO
perturbation theory) and the $Z_P$-ratios from the analysis of
sect.~\ref{combfit} (combined fit) or \ref{contfit} (fit in the
continuum).  The average light quark mass $m_{ud}({\overline{\mbox{\rm
MS}}},2{\rm GeV})$ from the fit in the continuum reads $m_{ud} =
3.43(9)(23)\, \textrm{MeV}$. From the combined fit we get $m_{ud} =
3.62(10)(23)$~MeV (using all the ensembles) and $m_{ud}
=3.49(13)(23)$~MeV (when excluding ensembles $B_6$ and $C_5$, {\it
i.e.} with the same set of ensembles as for the fit in the continuum).
When considering data at $\beta=3.9$ {\it only} (as in
sect.~\ref{singfit}), we obtain $m_{ud} = 3.62(13)(23)$~MeV. This
value is compatible with the one coming from the partially quenched
analysis of Ref.~\cite{ETM2}, $m_{ud} = 3.85(12)(40)\,\textrm{MeV}$.

The chiral quark condensate is obtained through $\langle \bar{q} q
\rangle({\overline{\mbox{\rm MS}}},2{\rm GeV}) = - (1/2) f_0^2
\widehat{B}_0({\overline{\mbox{\rm MS}}},2{\rm GeV})$.  From the fit
after the continuum extrapolation we get $|\langle \bar{q} q
\rangle|^{1/3} = 272(4)(7)\, \textrm{MeV}$, while from the combined
fit analysis we obtain $|\langle \bar{q} q \rangle|^{1/3} =
267(4)(7)\, \textrm{MeV}$ (and $|\langle \bar{q} q \rangle|^{1/3} =
270(4)(7)\, \textrm{MeV}$ when the ensembles $B_6$ and $C_5$ are not
used).  These results are in agreement with an independent estimate at
$\beta=3.9$ from the $\epsilon$ regime -- see Ref.~\cite{AS-PROC}.

\section{Scaling analysis of other hadronic observables}

Here we briefly report on the scaling behaviour of other hadronic
observables. Even if at $\beta=3.8$ preliminary data are available
only for the meson vector mass and the physical volumes at the various
lattice resolutions ($L \simeq 2.1$~fm for $\beta = 4.05, 3.9$ and $L
\simeq 2.4$~fm for $\beta = 3.8$) are only approximatively matched,
this overview suggests that when employing maximally twisted Wilson
quarks the cutoff effects on physical quantities are in general as
small as expected on the basis of O($a$) improvement.
\subsection{Nucleon and vector meson masses}
\begin{figure}[!t]
  \begin{center}
    \hspace*{-6.5cm} \includegraphics[scale=0.70,angle=0]{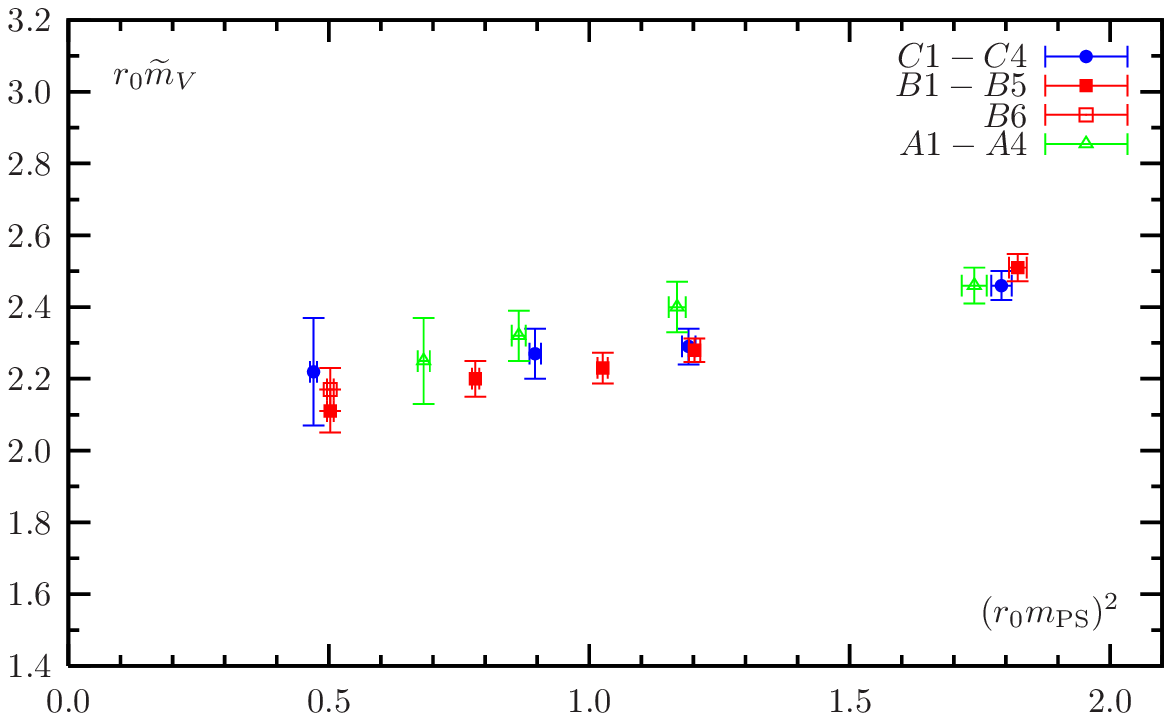}
  \end{center}
  \vspace*{6.05cm}
  \begin{center}
    \hspace*{0.5cm}  \includegraphics[scale=0.693,angle=0]{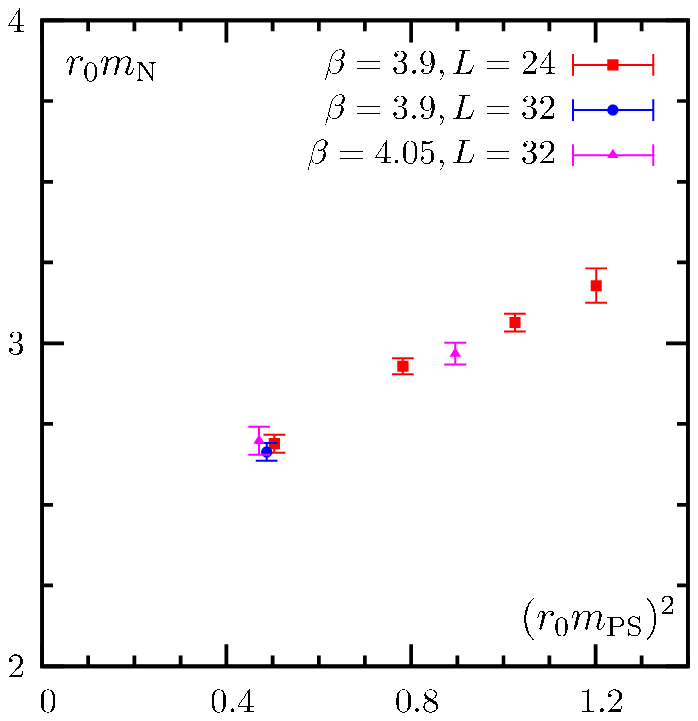}
    \vspace*{-6.9cm}
    \caption[]{$\widetilde{m}_{\rm V}r_0$ and $m_{\rm N}r_0$ vs.\
      $(m_{\rm PS}r_0)^2$ for different lattice resolutions and $L$ in
      the range $2.1 \div 2.4$~fm.}
    \label{FIG:V-N-mass}
  \end{center}
\end{figure}
\vspace*{-0.3cm}

In Fig.~\ref{FIG:V-N-mass}, we show our data (in units of $r_0$) for
the vector meson ``mass'', $\widetilde{m}_{\rm V}$, and the nucleon
mass, $m_{\rm N}$, as a function of $m_{\rm PS}^2$.  For the former
one (see Ref.~\cite{CM-PROC} for more details), the scaling appears to
be good, although within statistical errors that become of few
percents when $m_{\rm PS} \sim 300 \div 350$~MeV. In this case, we
remark that what we call the ``mass'' may differ from the actual
vector meson mass (the one that becomes $m_\rho$ as $m_{\rm PS} \to
m_\pi$), owing to the effect of virtual $\rho$--$\pi\pi$
mixing.\,\footnote{ The correction is estimated~\cite{CM-PROC} to be
almost irrelevant within our present statistical errors: at our most
chiral point, where $m_{\rm PS} \sim 300$~MeV, it amounts to
$\widetilde{m}_{\rm V} - m_{\rm V} \sim 0.05 m_{\rm V}$.}  Due to the
finite volume, the decay of the $\rho$-meson (at rest) into real
$\pi\pi$ states is instead forbidden, even when $2m_{\rm PS}< m_{\rm
V}$ (which roughly speaking happens in the region $(m_{\rm PS} r_0)^2
\lesssim 1$).
The data for the nucleon mass are also preliminary (see
Ref.~\cite{CA-PROC}): at $\beta=4.05$ only two data points are
available, which however appear to agree quite well with the data
points at $\beta=3.9$.

\subsection{Charmed observables in partially quenched setup}

\begin{figure}[!t]
\begin{center}
\subfigure[]{\includegraphics[width=.46\textwidth]{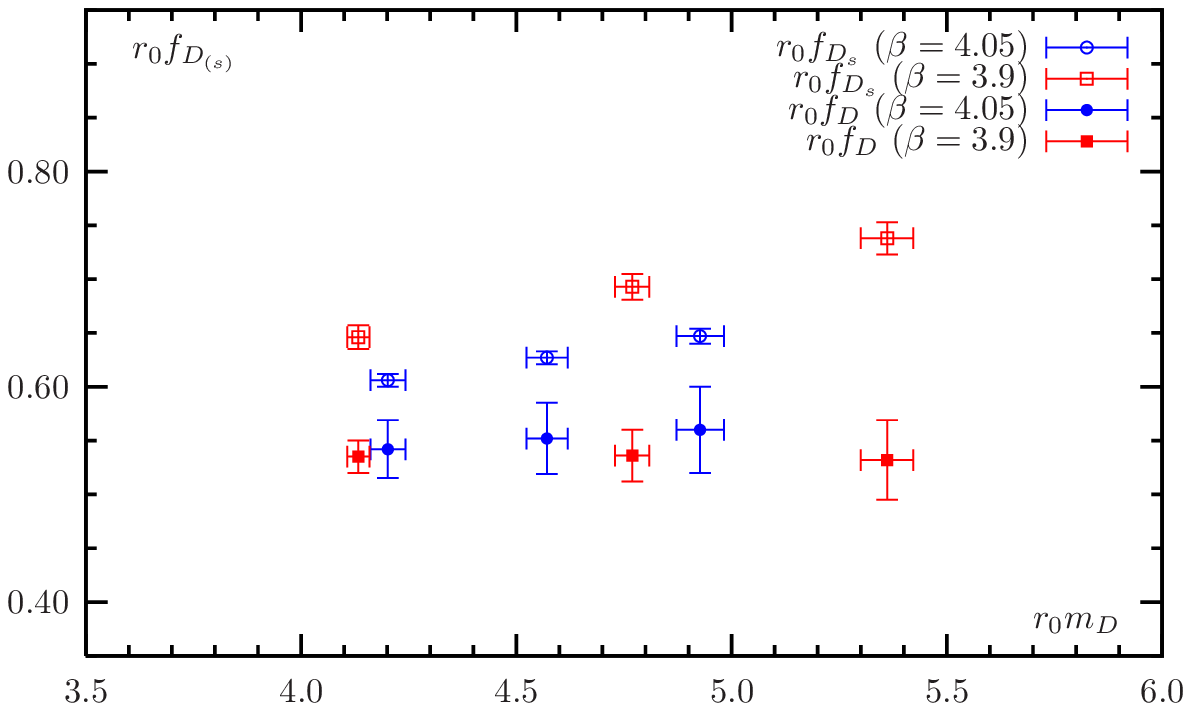}}
\hspace{0.8cm}
\subfigure[]{\includegraphics[width=.46\textwidth]{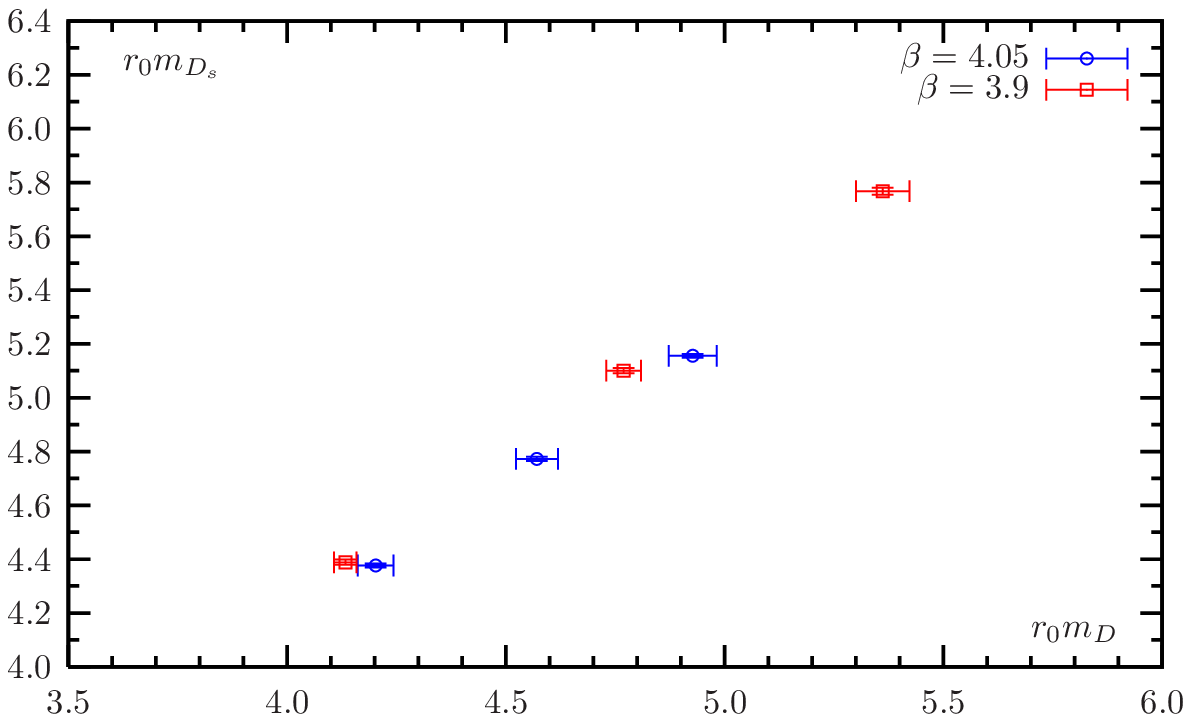}}
\vspace{-0.3cm}
\caption[]{$f_{\rm D}r_0$, $f_{\rm D_s}r_0$ (a) and~$m_{\rm D_s}r_0$
     (b) vs.\ $m_{\rm D}r_0$ for two lattice spacings ($\beta=4.05$
     and $3.9$) and $L \sim 2.1$~fm.}
\label{FIG:charmedobs}
\end{center}
\end{figure}

To compute charmed observables we adopt a partially quenched (PQ)
setup, which is detailed in Ref.~\cite{BB-PROC}.\,\footnote{The proof
of automatic O($a$) improvement given in Ref.~\cite{FR2} for a fully
unquenched mixed action framework goes through also in the PQ mixed
action setup of Ref.~\cite{BB-PROC}, where the quark masses of the
valence $s$ and $c$ quarks are taken finite and hence different from
those (infinite in the present case) of the corresponding sea quarks.}
With the notation of Ref.~\cite{FR2} for valence quarks, the
correlators for $D$ ($\bar{c}d$) and $D_s$ ($\bar{c}s$) charged mesons
are computed with Wilson parameters $r_d = r_s = -r_c =1$, in order to
have the same nice parametric scaling properties as in the charged
pion sector.  For the present considerations about scaling we choose
close-to-realistic renormalization conditions: $m_{\rm PS} r_0 =
0.7092$ and $\mu_{\tilde{s}}/\mu_{\tilde{c}} = 0.082$ (instead of the
realistic conditions $m_{\pi} r_0 \simeq 0.30$ and $\mu_{s}/\mu_{c}
\simeq 0.088$).  In Figs.~\ref{FIG:charmedobs}, $m_{\rm D}$ varies
with $\mu_{\tilde{c}}$ at fixed values of
$\mu_{\tilde{s}}/\mu_{\tilde{c}}$ and $m_{\rm PS} r_0$ ({\it i.e.}
$\mu_{ud}$).  The expected size of the dominating cutoff effects is
$(a\mu_{\tilde{c}})^2 \sim 0.1 $, since typically, $a\mu_{\tilde{s}}
\in [0.020, \, 0.033]$ and $a\mu_{\tilde{c}} \in [ 0.25, \, 0.40]$.
By comparing the preliminary results from $\beta=4.05$ and
$\beta=3.9$, we see that the observed scaling violations are not
large: from 1--2\% (hardly visible within statistical errors) for
$f_{\rm D}$ and $m_{\rm D_s}$ to 7--8\% for $f_{\rm D_s}$. Data from a
third lattice spacing, $\beta=3.8$, are currently being analysed and
may allow for an estimate of continuum limit results.

\section{Conclusions and outlook}

We have reported on the scaling properties of lattice QCD with $N_{\rm
f}=2$ maximally twisted Wilson quarks, in the unitary~\cite{CU-PROC}
as well as in a partially quenched setup~\cite{ETM2,BB-PROC}.  Very
good scaling properties are found in the light (charged) PS-meson
sector and also for various charmed PS-meson observables, as well as
in the vector meson and the parity-even nucleon channels, in agreement
with the expectation of automatic O($a$) improvement of physical
observables.  Based on these findings, we presented $\chi$PT-based
analyses suggesting that the determinations of the LEC's (among which
those relevant for the chiral condensate) and of the average $u,d$
quark mass that were obtained for one single lattice resolution in
Refs.~\cite{ETM1,ETM2} are likely to be close to their continuum limit
$N_{\rm f}=2$ QCD values.  Moreover, combining the results presented
here with those of Ref.~\cite{CA-PROC}, after appropriate chiral
extrapolations, one finds an estimate of the ratio $m_{\rm N}/f_{\rm
\pi}$ that is in good agreement with experiment.  The framework of
lattice QCD with maximally twisted Wilson quarks appears thus to offer
good prospects for reliable computations of many physical QCD
observables and weak matrix elements in the continuum limit.

\section*{Acknowledgements}

\noindent We thank all our ETM collaborators for many important
contributions to this work. In particular we are grateful to those who
directly provided us with specific results: C.~Alexandrou (nucleon
mass), B.~Blossier (charmed observables -- besides a crucial
contribution to the computation of $r_0$ at $\beta=3.8$), C.~Michael
(vector meson mass), V.~Lubicz and S.~Simula (renormalization
constants). This work was partially supported by the EU Contract
MRTN-CT-2006-035482 ``FLAVIAnet''.

\end{document}